\documentclass[twocolumn, pra, superscriptaddress]{revtex4-1}
\usepackage{amssymb}
\usepackage{amsmath}
\usepackage{epsfig}
\usepackage{color}
\usepackage{graphics, graphicx}
\usepackage{bbold}
\usepackage{psfrag}
\usepackage{mathcomp}
\usepackage{subfigure}
\usepackage{verbatim}
\usepackage{color}
\usepackage[colorlinks,citecolor=blue]{hyperref}
\def\cp#1{\mathbf{#1}}

\begin{document}
\title{Dissipation-facilitated molecules in a Fermi gas with non-Hermitian spin-orbit coupling}
\author{Lihong Zhou}
\affiliation{Beijing National Laboratory for Condensed Matter Physics, Institute of Physics, Chinese Academy of Sciences, Beijing 100190, China}
\author{Wei Yi}
\email{wyiz@ustc.edu.cn}
\affiliation{CAS Key Laboratory of Quantum Information, University of Science and Technology of China, Hefei 230026, China}
\affiliation{CAS Center For Excellence in Quantum Information and Quantum Physics, Hefei 230026, China}
\author{Xiaoling Cui}
\email{xlcui@iphy.ac.cn}
\affiliation{Beijing National Laboratory for Condensed Matter Physics, Institute of Physics, Chinese Academy of Sciences, Beijing 100190, China}
\affiliation{Songshan Lake Materials Laboratory , Dongguan, Guangdong 523808, China}
\date{\today}

\begin{abstract}
We study the impact of non-Hermiticity on the molecule formation in a two-component spin-orbit-coupled Fermi gas near a wide Feshbach resonance.
Under an experimentally feasible configuration where the two-photon Raman process is dissipative, the Raman-induced synthetic spin-orbit coupling acquires a complex strength.
Remarkably, dissipation of the system facilitates the formation and binding of molecules, which, despite their dissipative nature and finite life time, exist over a wider parameter regime than in the corresponding Hermitian system.
These dissipation-facilitated molecules can be probed by the inverse radio-frequency (r.f.) spectroscopy, provided the Raman lasers are blue-detuned to the excited state. The effects of dissipation manifest in the r.f. spectra as shifted peaks with broadened widths, which serve as a clear experimental signature.
Our results, readily observable in current cold-atom experiments, shed light on the fascinating interplay of non-Hermiticity and interaction in few- and many-body open quantum systems.
\end{abstract}

\maketitle

\section{Introduction}
Non-Hermitian physics have been intensively explored recently in a wide range of experimental systems such as optics, acoustics, and microwave cavities~\cite{review1,review2}.
Whereas most of these studies focus on single-particle physics, the interplay of non-Hermiticity and interaction, a key element in understanding non-Hermitian many-body quantum systems, is a more challenging but less explored subject. An ideal platform for this study is cold atomic gases, where the interaction is highly tunable through Feshbach resonances~\cite{FR}, and non-Hermiticity can be conveniently implemented via laser-induced one-body~\cite{Luo, Gadway,Gerbier,Ott1,Ott2,Ott3} or two-body~\cite{Takahashi1,Takahashi2,Durr} dissipation. Up to now, the existing experimental~\cite{Gerbier,Ott1,Ott2,Ott3,Takahashi1,Takahashi2,Durr} and theoretical~\cite{Ott4, Ott5, Wunner, Main, Konotop, Ueda1, Pan, Yu, Ueda2, zhaichen} studies have predominantly focused on properties of interacting bosons with dissipation, while the many-body physics of dissipative fermions have rarely been discussed, except for a limited number of theoretical works exploring the fermion superfluidity in various non-Hermitian settings~\cite{Ghatak, Zhou, Yamamoto}.
At this stage, it is highly desirable to search for readily accessible non-Hermitian systems, fermionic in particular, where the interplay of interaction and non-Hermiticity leads to non-trivial and experimentally detectable phenomena.

\begin{figure}[t]
\includegraphics[width=5cm]{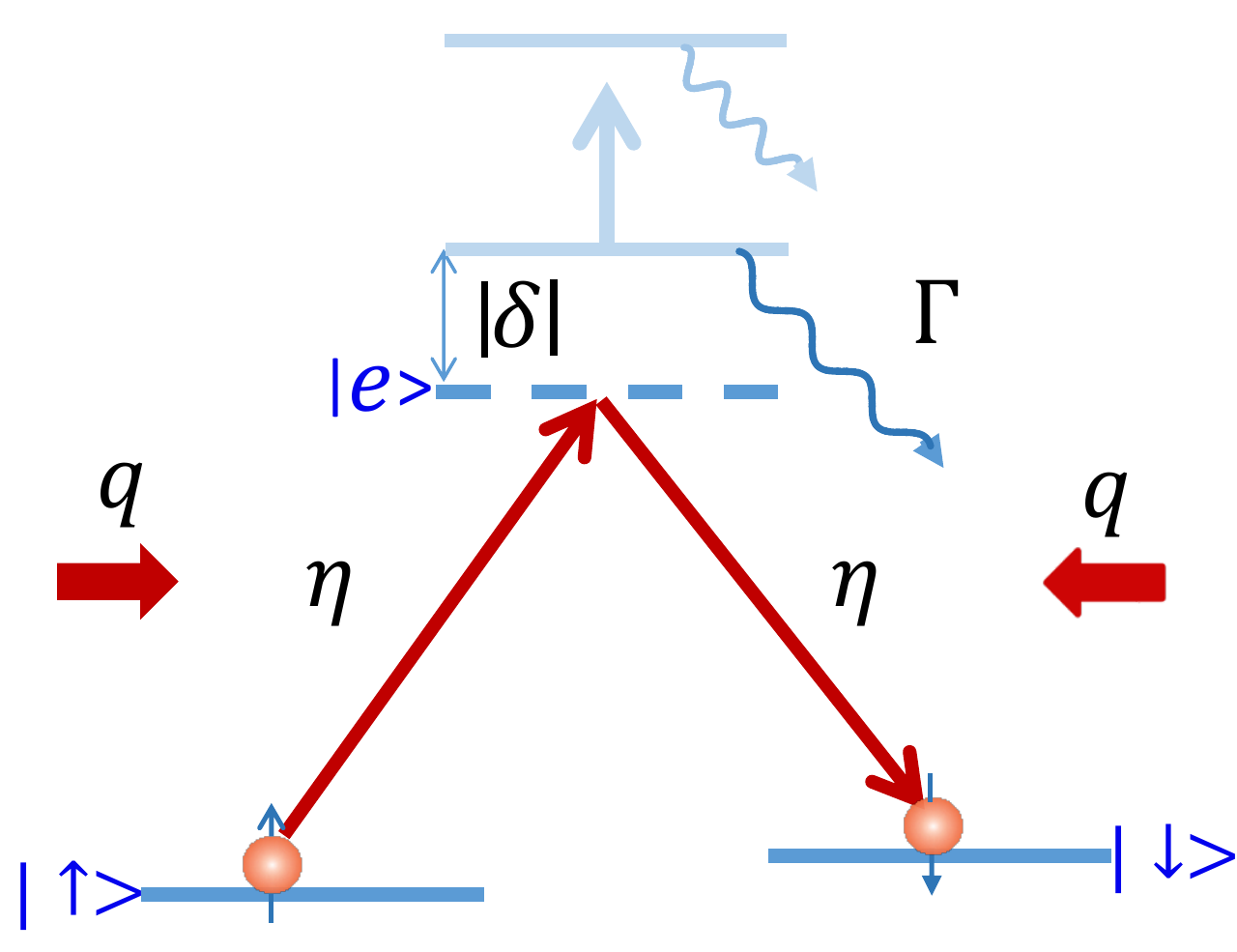}
\caption{(Color online) Schematics of the experimental setup. A two-component Fermi gas with hyperfine spin states $|\uparrow\rangle$ and $|\downarrow\rangle$ are subject to a non-Hermitian SOC through a dissipative Raman process. The parameters $q$, $\eta$ and $\delta$ respectively represent the transferred momentum, single-photon Rabi frequency and detuning. The tunable effective decay rate $\Gamma$ of the intermediate excited state ($|e\rangle$) contains contribution from both spontaneous decay and laser-induced decay. Here we show only Raman lasers with red detuning as an example.
} \label{fig_setup}
\end{figure}

In this work we propose an experimentally feasible scheme where the non-trivial effect of non-Hermitian spin-orbit coupling (SOC) on the molecule formation in a two-component Fermi gas can be probed by the widely used radio-frequency (r.f.) spectroscopy~\cite{rfreview}. We start from the two-photon Raman process that has been used to generate SOCs in Hermitian cold-atom systems~\cite{SOC_expt1,SOC_expt2, SOC_expt3, SOC_expt4, SOC_expt5}. To make the SOC non-Hermitian, we consider the case where
the intermediate excited state is subject to a controllable loss, which, for instance, can be induced by an additional laser, as illustrated in Fig.~\ref{fig_setup}. Such a loss channel makes the SOC strength complex-valued, with tunable real and imaginary parts.
Since the SOC does not commute with inter-atomic interactions, the resulting non-Hermitian system is expected to host non-trivial few- and many-body quantum phenomena.

Here we focus on the study of two-body bound states, or molecules, under the non-Hermitian SOC. We find that, in contrast to previously realized spin-selective dissipation~\cite{Luo, Gadway}, which do not affect the molecular binding of spin-singlet fermions~\cite{footnote}, the non-Hermitian SOC has significant impact on molecules.
Remarkably, the non-Hermitian SOC can greatly facilitate the molecule formation, in that it can induce new molecular branches or enhance the binding energy of existing molecules in a wide parameter regime. This is in distinct contrast to Hermitian systems where the molecule formation is suppressed by SOC~\cite{mol1,mol2,mol3}. Another important feature of these molecules is that their binding energies are generally complex-valued with negative imaginary parts, reflecting their dissipative nature with finite life time.
Nevertheless, we show that these molecules can still be experimentally probed via the inverse radio-frequency (r.f.) spectroscopy.
Signatures of non-Hermiticity manifest themselves as shifted spectral peaks and broadened widths in the r.f. spectrum.
Our results pave the way for experimental explorations of fundamental few-body physics in non-Hermitian fermionic systems, which are indispensable for the understanding of many-body phenomena therein.

The work is organized as follows. In Sec.~II, we derive the non-Hermitian effective Hamiltonian of the system starting from the Lindblad equation. We then study the single-particle physics in Sec.~III, before fully characterizing the dissipation-facilitated molecules in Sec.~IV. In Sec.~V, we show that signals of the dissipative molecules can be detected through r.f. spectroscopy. Finally, we summarize in Sec.~VI.

\section{Model}
We start by writing down the Lindblad master equation corresponding to the configuration in Fig.~\ref{fig_setup} ($\hbar=1$ throughout the paper)
\begin{align}
\frac{d}{dt}\rho&=-i\Big[H,\rho\Big]-\frac{1}{2}\Gamma\left(S^\dag S\rho+\rho S^\dag S-2S\rho S^{\dag}\right)\nonumber\\
&=-i\left(H_{\rm eff}\rho-\rho H_{\rm eff}^\dag\right)+\Gamma S\rho S^\dag.\label{eq:lindblad}
\end{align}
Here the single-particle Hamiltonian $H$ for the internal degrees of freedom is given by
\begin{align}
H=-\delta |e\rangle\langle e|+\eta \left(e^{iqx}|e \rangle\langle \uparrow|+e^{-iqx}|e\rangle\langle \downarrow|+h.c.\right),
\end{align}
where $|e\rangle$ is the excited state, $|\sigma\rangle$ ($\sigma=\uparrow,\downarrow$) are the ground hyperfine spin states, and  $q$ is the transferred momentum.
Dissipation of the system originates from either the spontaneous decay of state $|e\rangle$, or the laser-induced decay of $|e\rangle$ to a third state, or both. These loss processes can be described by the quantum jump term in Eq.\ref{eq:lindblad}, where the quantum jump operator is given by $S=|r\rangle\langle e|$ ($|r\rangle$ is the reservoir state) and $\Gamma$ is the overall decay rate of $|e\rangle$.  We also assume the decay of $|e\rangle$ does not end up in states $|\sigma\rangle$. This can be achieved, for example, by coupling the excited state $|e\rangle$ to states which are not trapped, and are immediately lost from the system.

Adiabatically eliminating the excited state $|e\rangle$, we derive the equations of motion for the remaining density-matrix elements
\begin{align}
\frac{d}{dt}\tilde{\rho}_{\uparrow\uparrow}&=-i(\Omega-\Omega^\ast)\tilde{\rho}_{\uparrow\uparrow}-i\Omega e^{-i2qx}\tilde{\rho}_{\downarrow\uparrow}+i\Omega^\ast e^{i2qx}\tilde{\rho}_{\uparrow\downarrow},\label{eq:motion1}\\
\frac{d}{dt}\tilde{\rho}_{\downarrow\downarrow}&=-i(\Omega-\Omega^\ast)\tilde{\rho}_{\downarrow\downarrow}-i\Omega e^{i2qx}\tilde{\rho}_{\uparrow\downarrow}+i\Omega^\ast e^{-i2qx} \tilde{\rho}_{\downarrow\uparrow},\label{eq:motion2}\\
\frac{d}{dt}\tilde{\rho}_{\uparrow\downarrow}&=-i(\Omega-\Omega^\ast)\tilde{\rho}_{\uparrow\downarrow}-i\Omega e^{-i2qx}\tilde{\rho}_{\downarrow\downarrow}+i\Omega^\ast e^{i2qx}\tilde{\rho}_{\uparrow\uparrow},\label{eq:motion3}\\
\frac{d}{dt}\tilde{\rho}_{rr}&=i(\Omega-\Omega^\ast)(\tilde{\rho}_{\uparrow\uparrow}+\tilde{\rho}_{\downarrow\downarrow}),\label{eq:motion6}
\end{align}
where $\Omega=\eta^2/(\delta+i \Gamma/2)$, and $\tilde{\rho}$ denotes the density matrix in the subspace spanned by the states $\{|\uparrow\rangle,|\downarrow\rangle,|r\rangle \}$. Equations of motion (\ref{eq:motion1})(\ref{eq:motion2})(\ref{eq:motion3})(\ref{eq:motion6}) can be rearranged into a compact Lindblad form
\begin{align}
\frac{d}{dt}\tilde{\rho}&=-i\left(H_{\rm SOC}\tilde{\rho}-\tilde{\rho} H_{\rm SOC}^\dag\right)+\gamma\sum_{i=1,2} L_i\tilde{\rho} L_i^\dag,\label{eq:lindbladeff}
\end{align}
where the quantum jump operators $L_1=|r\rangle\langle \uparrow|$ and $L_2=|r\rangle\langle \downarrow|$, $\gamma=-2\text{Im}(\Omega)$,
and the non-Hermitian effective Hamiltonian is given by
\begin{align}
H_{\rm SOC}=\Omega |\uparrow\rangle\langle\uparrow|+\Omega|\downarrow\rangle\langle\downarrow|+\Omega\left( e^{-i2qx}|\uparrow\rangle\langle\downarrow|+H.c.\right).
\end{align}
Note that above Hamiltonian also includes a complex Stark-shift term (of strength $\Omega$) for each spin, which ensures the dissipative nature of all eigenstates of the system.

Time evolution of the system under Eq.~(\ref{eq:lindbladeff}) can be described as non-Hermitian dynamics driven by the non-Hermitian Hamiltonian $H_{\rm SOC}$, which is further interrupted by quantum jumps, dictated by the terms $\gamma L_i\rho L_i^\dag$ ($i=1,2$). For the short-time dynamics within the timescale $t\ll 1/\gamma$, the impact of quantum-jump terms is insignificant and the system dynamics is dominantly governed by  the non-Hermitian effective Hamiltonian $H_{\rm SOC}$. The physical origin of this relation is that the number loss of our system comes directly from the dissipative spin-orbit coupling (SOC) in the two-photon Raman process, therefore the imaginary SOC strength determines the quantum jump (or number decay) of such a system.
Note that in the far detuning regime $\eta\ll \sqrt{\delta^2+(\Gamma/2)^2}$, the excited state $|e\rangle$ is barely populated, giving rise to effective parameters such as $\Omega$ and $\gamma$ which govern the dynamics within the ground-state manifold $\{|\uparrow\rangle,|\downarrow\rangle\}$. The applicability of the non-Hermitian Hamiltonian is therefore restricted by $\gamma$ rather than $\Gamma$ in Eq.~(\ref{eq:lindblad}). It is also understood that the Markovian approximation is satisfied, with the bath correlation time much shorter than all relevant time scales in our study.


Combining $H_{\rm SOC}$ with the single-particle kinetic energy, we arrive at the following effective single-particle Hamiltonian
\begin{align}
H_0=\int d{\bf r} \psi^{\dag}({\bf r})\left[ (-\frac{\nabla^2}{2m}+\Omega)+\Omega(S_{+}e^{-i2qx}+H.c.) \right]\psi({\bf r}). \label{h0_single}
\end{align}
Here $\psi({\bf r})=[\psi_{\uparrow}({\bf r}),\psi_{\downarrow}({\bf r})]^T$, and $\psi_{\sigma}$ is the annihilation field operator for the state $|\sigma\rangle$; the operator $S_+$ ($S_-$) converts the spin-$\downarrow$ to spin-$\uparrow$ (spin-$\uparrow$ to spin-$\downarrow$). For the convenience of discussion, we rewrite the complex SOC strength $\Omega$ as
\begin{equation}
\Omega=\frac{\Omega_0}{1\pm i\tilde{\Gamma}}, \label{soc}
\end{equation}
where $+(-)$ corresponds to the blue (red) single-photon detuning with $\delta >0$ ($\delta<0$), $\Omega_0=\eta^2/\delta$ is the Raman-induced SOC strength at zero dissipation, and $\tilde{\Gamma}=\Gamma/(2|\delta|)$ is the dimensionless dissipation strength.

At short times ($t\ll 1/\gamma$), one may then identify the lowest real part of the eigen spectrum as the effective ground-state energy and the imaginary part as its decay rate~\cite{Zhou, Yamamoto}. We note that this is in contrast to previous studies on many-body steady states at long times in open systems~\cite{dissbec1,dissbec2}.

After a U(1) transformation of the field operators, $\psi_{\sigma}({\bf r})=e^{is_{\sigma}qx}\bar{\psi}_{\sigma}({\bf r})$ ($s_{\uparrow}=-1,\ s_{\downarrow}=1$), $H_0$ can be written in momentum space as $H_0=\sum_{\cp k} \bar{\psi}^{\dag}_{\cp k}h_0({\cp k})\bar{\psi}_{\cp k}$, with the transformed basis $\bar{\psi}_{\cp k}=[\bar{\psi}_{{\cp k}\uparrow},\bar{\psi}_{{\cp k}\downarrow}]^T$, and
\begin{equation}
h_0(\cp{k})=\frac{1}{2m}({\bf k}-q\sigma_z{\bf e}_x)^2+\Omega \sigma_x+\Omega I. \label{h0}
\end{equation}
Here $\sigma_{\alpha}$ ($\alpha=x,y,z$) is Pauli matrix and $I$ is $2\times 2$ identity matrix. The corresponding single-particle dispersion is
\begin{equation}
\xi_{\cp k\pm}=\epsilon_{\cp k}+E_q+\Omega\pm\sqrt{(qk_{x}/m)^2+\Omega^2}, \label{E_single}
\end{equation}
where $E_q=q^2/(2m)$, $\epsilon_{\cp k}={\cp k}^2/(2m)$, and $\pm$ indicates different helicity branches.

The total Hamiltonian is then given by $H=H_0+U_{\rm int}$, with the interaction part given by
\begin{equation}
U_{\text{int}}=U\sum_{{\cp Q},{\cp k},{\cp k'}} \bar{\psi}^{\dag}_{{\cp k}\uparrow} \bar{\psi}^{\dag}_{{\cp Q}-{\cp k}\ \downarrow}\bar{\psi}_{{\cp Q}-{\cp k'}\ \downarrow}\bar{\psi}_{{\cp k'}\uparrow}, \label{U}
\end{equation}
where the bare interaction $U$ is related to the $s$-wave scattering length $a_s$ via $1/U=m/(4\pi a_s)-1/V\sum_{\cp k} m/k^2$. Note that this renormalization relation is not changed by the presence of SOC that is linearly proportional to the momentum~\cite{Cui_RG}.

\begin{figure}[t]
\includegraphics[width=9cm]{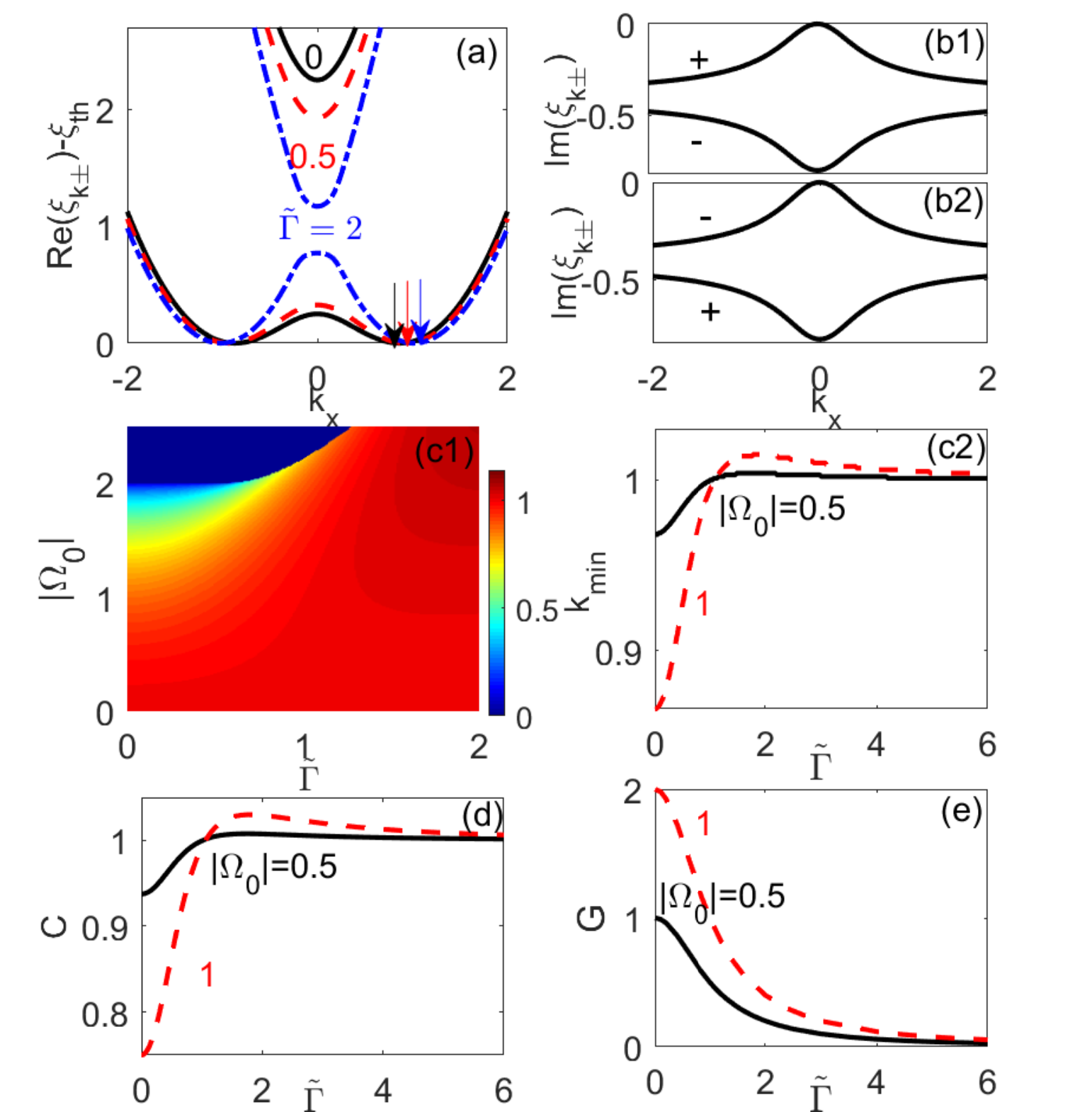}
\caption{(Color online) Single-particle physics modified by dissipation. (a) $\text{Re}(\xi_{\cp k \pm})$ (shifted by $\xi_{\rm th}$) along $k_x$ for different $\tilde{\Gamma}$ at a fixed $|\Omega_0|=1$. The arrows mark the locations ($ k_{\rm min}$) of the energy minimum. (b1)(b2) $\text{Im}(\xi_{\cp k \pm})$  for red- (b1) and blue-detuned (b2) Raman lasers at $|\Omega_0|=1$ and $\tilde{\Gamma}=0.5$. 
The helicity indices ($+/-$) are marked on the curves accordingly. (c1) Contour plot of $k_{\rm min}$ (encoded by the color bar) in the $(|\Omega_0|, \tilde{\Gamma})$ plane. (c2) $k_{\rm min}$ as a function of $\tilde{\Gamma}$ for given $|\Omega_0|$. (d) Coupling constant $C$ for two lower-helicity fermions at ${\cp k}_{m}$ and $-{\cp k}_{m}$
as a function of $\tilde{\Gamma}$. Here ${\cp k}_{m}=(k_{\rm min}, 0, 0)$. (e) Real energy gap $G$ at ${\cp k}=(0, 0, 0)$. In all figures, we take $q$ and $E_q$ as the units of momentum and energy, respectively.
 } \label{fig_single}
\end{figure}

\section{Single-particle physics}
We first study the impact of dissipation on the single-particle dispersion $\xi_{\cp k \pm}$. 
As shown in Fig.~\ref{fig_single}(a), $\text{Re}(\xi_{\cp k \pm})$ are identical for red- and blue-detuned lasers, if both are shifted by the corresponding threshold energy $\xi_{\rm th}=\text{Re}(\xi_{\cp k_m-})$. Here ${\cp k}_m=(k_{\rm min}, 0,0)$, and $k_{\rm min}$ is the location of minimum $\text{Re}(\xi_{\cp k -})$ along $k_x$. For $\text{Im}(\xi_{\cp k \pm})$, however, the red- and blue-detuned cases are different [see Fig.~\ref{fig_single}(b1)(b2)]. In either case, $\text{Im}(\xi_{{\cp k}\pm})<0$, indicating a finite life time for single particle.

For our later discussion of molecule formation, it is helpful to highlight several key properties of the single-particle dispersion.
First, it is found that $k_{\rm min}$ varies non-monotonically  with $\tilde{\Gamma}$: it increases for small $\tilde{\Gamma}$ and decreases for large $\tilde{\Gamma}$ [see Figs.~\ref{fig_single}(c1)(c2)]. Second, we examine the coupling constant between the lowest-energy states in the helicity basis, which is proportional to the following expectation value
\begin{equation}
C=\langle -{\cp k}^L_{m-};{\cp k}^L_{m-}| \sum_{\cp k\cp k'}\bar{\psi}^{\dag}_{{\cp k}{\uparrow}} \bar{\psi}^{\dag}_{-{\cp k}'\downarrow}\bar{\psi}_{-{\cp k}'\downarrow}\bar{\psi}_{{\cp k}\uparrow}|{\cp k}^R_{m-};-{\cp k}^R_{m-} \rangle, \label{C}
\end{equation}
with $|{\cp k}^{R/L}_{\mu}\rangle$ denoting the right/left eigenvector satisfying $h_0|{\cp k}^R_{\mu}\rangle=\xi_{{\cp k}\mu}|{\cp k}^R_{\mu}\rangle$ and $h_0^\dagger|{\cp k}^L_{\mu}\rangle=\xi^*_{{\cp k}\mu}|{\cp k}^L_{\mu}\rangle$.
It is found that $C$ also varies non-monotonically with $\tilde{\Gamma}$ [see Fig.~\ref{fig_single}(d)]. Furthermore, in Fig.~\ref{fig_single}(e), we show the energy gap $G=\text{Re}(\xi_{0 +})-\text{Re}(\xi_{0 -})$, which decays monotonically with $\tilde{\Gamma}$.

The non-monotonic behavior of $k_{\rm min}$ and $C$ can be understood through perturbation theory in either the small- or large-$\tilde{\Gamma}$ limit (see Appendix \ref{app_A}). In the large-$\tilde{\Gamma}$ limit, the SOC strength decays as $\Omega\sim i\Omega_0/\tilde{\Gamma}$ and  Eq.~(\ref{h0}) is reduced to the Hamiltonian studied in Ref.~\cite{Zhou}. The reduction of SOC strength in this limit is a direct consequence of the quantum Zeno effect, as observed previously in dissipative atomic systems~\cite{Ott1,Durr}.
An important implication from the non-monotonic behavior is that non-Hermitian SOC would achieve its strongest effect at an intermediate strength $\tilde{\Gamma}\sim 1$, which we confirm in the molecule calculations below.

\section{Dissipation-facilitated molecules}
The molecular state in our interacting non-Hermitian system
satisfies the Lippman-Schwinger equation
\begin{equation}
|\Phi^R \rangle=G_0 U_{\text{int}} |\Phi^R\rangle.
\end{equation}
Here $G_0=(E_2-H_0+i0^+)^{-1}$ is the non-interacting Green's function at the two-body energy $E_2=2\xi_{\rm th}+E_b$, where $E_b$ is the two-body binding energy with $\text{Im}(E_2)=\text{Im}(E_b)$.
Since $U_{\text{int}}$ only acts on the spin-singlet state $|S=0\rangle=\frac{|\uparrow \downarrow\rangle-|\downarrow \uparrow \rangle}{\sqrt{2}}$, we arrive at the following equation for the two-body state
\begin{equation}
 \frac{1}{U}=\langle S=0| G_0(0,0) |S=0\rangle, \label{2body}
\end{equation}
where the Green's function can be expanded as
\begin{align}
G_0({\cp r},{\cp r'})=\frac{1}{2}\sum_{{\cp k};\mu\nu} \frac{\langle {\cp r}|{\cp k}^R_{\mu};-{\cp k}^R_{\nu}\rangle \langle-{\cp k}^L_{\nu};{\cp k}^L_{\mu}|{\cp r'}\rangle}{\langle {\cp k}^L_{\mu}|{\cp k}^R_{\mu}\rangle \langle -{\cp k}^L_{\nu}|-{\cp k}^R_{\nu}\rangle (E_2-\xi_{{\cp k}\mu}-\xi_{-{\cp k}\nu})} .  \label{G}
\end{align}
Different from the Hermitian case, Eq.(\ref{2body}) leads to two coupled equations corresponding to the real and imaginary parts of the equation, which can be solved for the complex binding energy $E_b$.

\begin{figure}[t]
\includegraphics[width=9cm]{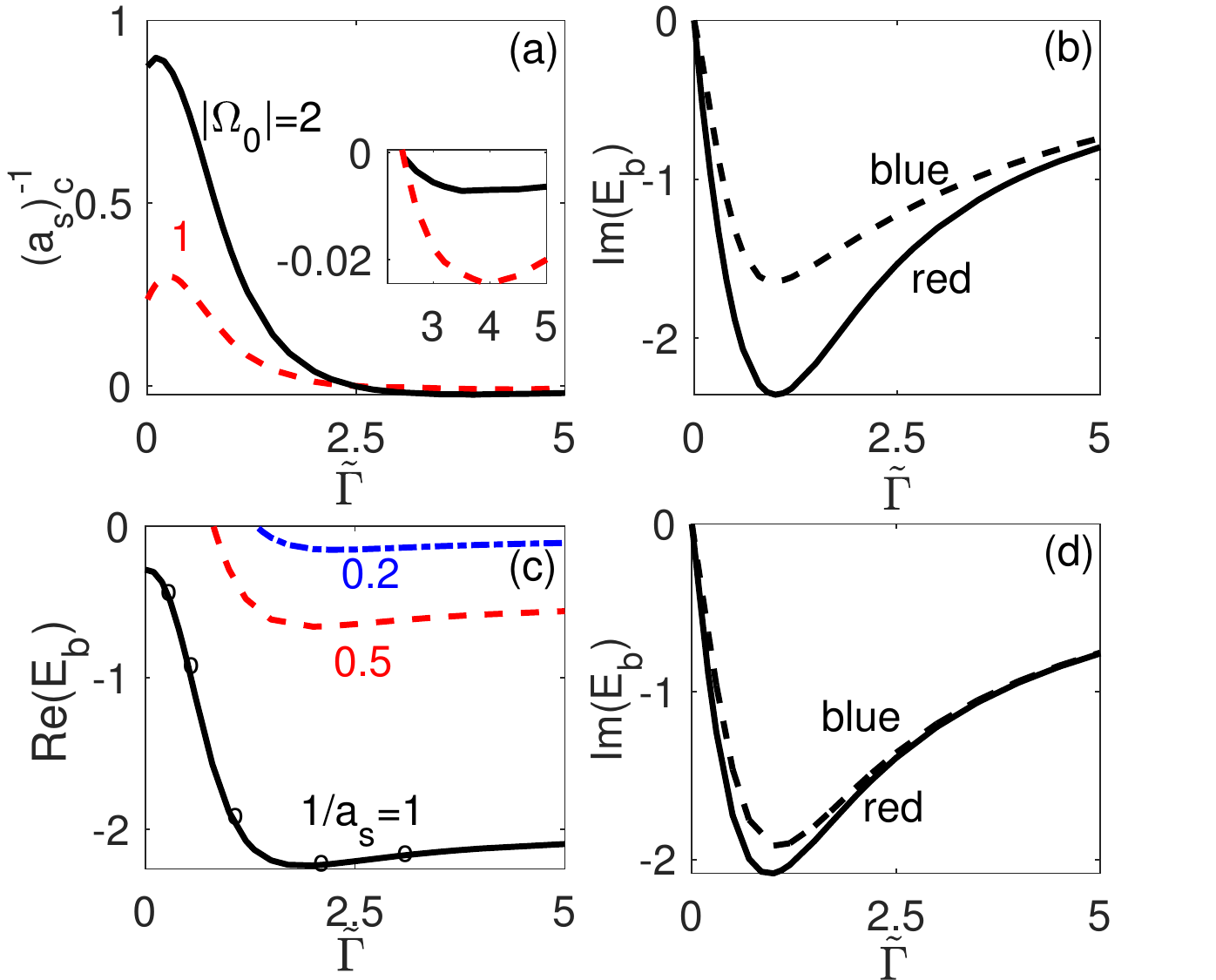}
\caption{(Color online) Dissipation-facilitated molecules.
(a) Critical coupling $(1/a_s)_c$ of two-body bound state as a function of $\tilde{\Gamma}$ for fixed $|\Omega_0|$. (Inset) An enlarged view of the regime where $(1/a_s)_c$ turns negative. (b) $\text{Im}(E_b)$ at the critical coupling as a function of $\tilde{\Gamma}$ for the red- ($\Omega_0=-2$) and blue-detuned ($\Omega_0=2$) Raman processes. (c) $\text{Re}(E_b)$ as functions of $\tilde{\Gamma}$ for several fixed couplings $1/a_s=0.2, \ 0.5,\ 1$, with $|\Omega_0|=2$. (d) $\text{Im}(E_b)$ at a fixed coupling strength $1/a_s=1$ for the red- ($\Omega_0=-2$) and blue-detuned ($\Omega_0=2$) Raman processes. We have taken $q$ and $E_q$ as the units of momentum and energy, respectively.
} \label{fig_Eb}
\end{figure}

Fig.~\ref{fig_Eb} shows typical results of the molecular solution. To demonstrate the effect of dissipation, in Fig.~\ref{fig_Eb}(a), we examine the critical coupling $(1/a_s)_c$ to support a two-body bound state as a function of $\tilde{\Gamma}$, obtained by setting $\text{Re}(E_b)=0$ in Eq.~(\ref{2body}). We can see that apart from a narrow region at very small $\tilde{\Gamma}$, $(1/a_s)_c$ is greatly reduced with increasing $\tilde{\Gamma}$, even to negative values for certain $\tilde{\Gamma}$ [see inset of Fig.~\ref{fig_Eb}(a)]. Therefore, molecules can form on the Bardeen-Cooper-Schieffer(BCS) side of the Feshbach resonance, in contrast to the case under one-dimensional Hermitian SOC, where molecules only survive on the Bose-Einstein condensate(BEC) side with positive $a_s$~\cite{mol1,mol2,mol3}.
The presence of dissipation-facilitated molecules is further confirmed in Fig~.\ref{fig_Eb}(c): for a fixed $1/(qa_s)=0.5$ (or $0.2$), a new molecular branch emerges at $\tilde{\Gamma}\geqslant0.82$ (or $\geqslant1.35$); for a stronger interaction $1/(qa_s)=1$, the molecule binding energy is enhanced by nearly an order of magnitude when increasing $\tilde{\Gamma}$ from $0$ to $2$.  At sufficiently large $\tilde{\Gamma}$, molecular energy saturates at $\text{Re}(E_b)=-1/(ma_s^2)$, as quantum Zeno effects become dominant.

The mechanism of dissipation-facilitated molecules is closely related to the single-particle physics discussed previously: the enhanced low-energy coupling constant [Fig.~\ref{fig_single}(d)] and the reduced energy gap [Fig.~\ref{fig_single}(e)] under dissipation. These factors make the pairwise scattering of fermions much easier in the low-energy subspace, giving rise to enhanced molecule formation. Moreover, such enhancement is most dramatic at intermediate dissipation with $\tilde{\Gamma}\sim 1$. This, again, is consistent with the non-monotonic behavior we show in Fig.~\ref{fig_single}.

Here we emphasize that the type of Raman detuning (red- or blue-detuned), which determines the sign of $\Omega_0$, does not alter the results of $(1/a_s)_c$ and $\text{Re}(E_b)$, but it does change the values of $\text{Im}(E_b)$ as shown in Fig.\ref{fig_Eb}(b)(d). For both red- and blue-detuned Raman lasers, we have $\text{Im}(E_b)<0$, suggesting that the molecules are all dissipative and have finite life time. 
We have checked that the imaginary part of binding energy are generally of the same order as the imaginary part of the SOC strength, i.e., $|\text{Im}(E_b)| \sim 2|\text{Im}(\Omega)|$, which can be attributed to the fact that $E_b$ is closely related to the complex single-particle spectrum in the presence of a complex SOC.

\section{Detection}
We now turn to the experimental detection of dissipation-facilitated molecules using r.f. spectroscopy. Consider a third hyperfine state $|1\rangle$, which has no interaction or SOC with $|\uparrow\rangle$ and $|\downarrow\rangle$, but can be coupled to $|\downarrow\rangle$ by a r.f. field with the Hamiltonian
\begin{equation}
H_{\rm rf}=\Omega_{\rm rf}\int d{\bf r} \left( e^{-i\omega t+iqx} \psi_1^{\dag}({\bf r},t) \bar{\psi}_{\downarrow}({\bf r},t) +H.c. \right).
\end{equation}
We focus on two different types of commonly used r.f. measurements in cold atoms: the direct r.f. spectroscopy, where the r.f. field breaks preformed molecules and transfers atoms from the $|\downarrow\rangle$ state to an empty $|1\rangle$ state; and the inverse r.f. spectroscopy, where atoms are initialized in $|1\rangle$ and are transferred to $|\downarrow\rangle$ state by the r.f. field to form the molecular state. Both types of measurements have been successfully implemented in cold Fermi gases under Hermitian SOC~\cite{SOC_expt4, Zhang_expt}.


In the background of the bi-orthogonal basis, the right and left states in the Schr\"odinger picture are
\begin{align}
&|\phi_s^R(t)\rangle=e^{-i(H+H_{rf})t}|\phi^R(0)\rangle,\\
&|\phi_s^L(t)\rangle=e^{-i(H+H_{rf})^\dagger t}|\phi^L(0)\rangle,
\end{align}
 here $|\phi^{R}(0)\rangle$ ($|\phi^{L}(0)\rangle$) is the initial state at $t=0$, which is usually an eigenstate of $H$ ($H^{\dag}$) with eigen-energy $E$ ($E^*$); $H_{rf}$ is the external field as the perturbation. The Hamiltonian of the system (without perturbation) is $H=H_0+U_{int}+H_1$,  and $H_1=\int d{\cp{r}}\psi_1^\dagger(\cp{r})(-\frac{\nabla^2}{2m})\psi_1(\cp{r})$.

The right and left states in the interaction picture are defined as
 \begin{align}
&|\phi^R(t)\rangle=e^{iHt}|\phi_s^R(t)\rangle,\label{interactionr}\\
&|\phi^L(t)\rangle=e^{iH^\dagger t}|\phi_s^L(t)\rangle,
\label{interactionl}
\end{align}
According to Eqs.~(\ref{interactionr},\ref{interactionl}), the expectation values of annihilation operator ($\psi_\alpha$) and creation operator $\psi_\alpha^\dagger(\alpha=\updownarrow,1)$ in the Schr\"odinger picture are
\begin{align}
&\langle \phi_s^L(t)|\psi_\alpha|\psi_s^R(t)\rangle=\langle \phi^L(t)|e^{iHt}\psi_\alpha e^{-iHt}|\phi^R(t)\rangle,\\
&\langle \phi_s^L(t)|\psi_\alpha^\dagger|\phi_s^R(t)\rangle=\langle \phi^L(t)|e^{iHt}\psi_\alpha^\dagger e^{-iHt}|\phi^R(t)\rangle,
\end{align}
so the field operators in the interaction picture can be defined as
\begin{align}
&\psi_{\alpha}(t)=e^{iHt}\psi_\alpha e^{-iHt},\\
&\psi_{\alpha}^\dagger(t)=e^{iHt}\psi_\alpha^\dagger e^{-iHt}.
\end{align}
Note that since we have $H\neq H^{\dag}$ for the non-Hermitian system,  the creation and annihilation field operators in the interaction picture $(\psi_{\alpha}(t),\psi^{\dag}_{\alpha}(t))$ are no longer conjugate to each other. This is the unique property of non-Hermitian systems.

According to the linear response theory, we consider the transition rate caused by the r.f. field that is associated with the retarded spin-flip correlation function
\begin{align}
D(t,t')&=-i\theta(t-t')\int d\cp{r}d\cp{r'}{}_L\langle [\psi_1^\dagger(\cp{r},t)\psi_{\downarrow}(\cp{r},t),\nonumber\\
&\psi_{\downarrow}^\dagger(\cp{r'},t')\psi_1(\cp{r'},t')]\rangle_R.
\label{retarded}
\end{align}
Its Fourier transform in the frequency space is
\begin{equation}
D(i\omega)=\frac{1}{\beta}\sum_{\cp{k}}\sum_n G_{\downarrow\downarrow}({\cp k},i\omega_n)G_{11}({\cp k}+q{\cp e}_x,i\omega_n\pm i\omega),
\end{equation}
where the sign $+(-)$ corresponds to the direct (inverse) r.f. process, $\omega$ and $\omega_n$ are both fermonic Matsubara frequencies, $G_{11}({\cp k},i\omega_n)=(i\omega_n-\epsilon_{\cp k})^{-1}$ is the Green's function for state $|1\rangle$, and $G_{\downarrow\downarrow}$ is the Green's function for state $|\downarrow\rangle$, which, in view of the molecular state, can be written as
\begin{equation}
 G_{\downarrow\downarrow}(\cp{k},i\omega_n)=\sum_{\lambda=\pm} \frac{\langle -{\cp k}^L_{\lambda}|\bar{\psi}_{{\cp k} \downarrow}|\Phi^R\rangle\langle\Phi^L|
 \bar{\psi}^{\dag}_{{\cp k}\downarrow}| -{\cp k}^R_{\lambda}\rangle}{i\omega_n-(E_2-\xi_{-{\cp k}\lambda})}.
 \end{equation}
Here $|\Phi^R\rangle$ ($|\Phi^L\rangle$) is the right (left) eigenvector for the molecular state with eigenenergy $E_2$ ($E_2^*$).

The transition rate is then given by
\begin{equation}
R(\omega)=-{\rm Im} D(i\omega\rightarrow \omega+i0^+).
\end{equation}
For a direct r.f. process, it leads to
\begin{equation}
R_d(\omega)=-\sum_{\cp{k}}\sum_{\lambda=\pm} {\rm Im} \left[ \frac{\langle -{\cp k}^L_{\lambda}|\bar{\psi}_{{\cp k} \downarrow}|\Phi^R\rangle \langle\Phi^L|
 \bar{\psi}^{\dag}_{{\cp k}\downarrow}| -{\cp k}^R_{\lambda}\rangle}{\omega+ (E_2-\xi_{-{\cp k}\lambda}-\epsilon_{{\cp k}+q{\cp e}_x})+i0^+} \right]; \label{rate_direct}
\end{equation}
For the inverse r.f. process, it leads to
\begin{align}
R_i(\omega)&=-\sum_{\cp{k}}\sum_{\lambda=\pm}n_F(\epsilon_{{\cp k}+q{\cp e}_x})
\nonumber\\
&\times {\rm Im} \left[ \frac{\langle -{\cp k}^L_{\lambda}|\bar{\psi}_{{\cp k} \downarrow}|\Phi^R\rangle \langle\Phi^L|
 \bar{\psi}^{\dag}_{{\cp k}\downarrow}| -{\cp k}^R_{\lambda}\rangle}{\omega- (E_2-\xi_{-{\cp k}\lambda}-\epsilon_{{\cp k}+q{\cp e}_x})+i0^+} \right] \label{rate_inverse}
\end{align}

In the Hermitian case with $\tilde{\Gamma}=0$, Eqs.~(\ref{rate_direct})(\ref{rate_inverse}) can be reduced to the Fermi's golden rule as applied to Hermitian SOC systems (see Appendix \ref{app_B}), which guarantees a positive-definite transition rate~\cite{Hu,Zhang_expt}. However, for a non-Hermitian system with $\tilde{\Gamma}>0$, due to the complex eigen-energies and eigenstates involved in Eqs.~(\ref{rate_direct})(\ref{rate_inverse}), the Fermi's golden rule breaks down in general. Accordingly, a crucial question is that, whether the transition rate can still be positive (rather than negative) to allow the experimental detection of molecules?

By noting that the molecules undergo a dissociation (association) process in the direct (inverse) r.f. spectroscopy, we recognize that important insights can be gained by comparing the lifetime of the molecule $\tau_m$ (as inferred from its imaginary energy) and that of unbound fermions $\tau_f$. Specifically, there are two scenarios:

(i) In the direct r.f. spectroscopy, the molecule is formed before the r.f. excitation. It follows that, to ensure a positive transition rate,  the molecule should be relatively more {\it stable} compared to unbound fermions after the r.f. excitation , i.e., $\tau_m>\tau_f$. Mathematically, this is equivalent to requiring $-\text{Im}(E_b)<-\text{Im}(\xi_{\cp k \lambda})~$\cite{footnote_helicity}, such that the denominator in the summand of Eq.~(\ref{rate_direct}) has a positive imaginary part.

(ii) In the inverse r.f. spectroscopy, the molecule is formed after the r.f. excitation, then a positive transition rate would require  $\tau_m<\tau_f$. This is equivalent to $-\text{Im}(E_b)>-\text{Im}(\xi_{\cp k \lambda})$, such that the denominator in the summand of Eq.~(\ref{rate_inverse}) again has a positive imaginary part.

\begin{figure}[t]
\includegraphics[width=8.5cm]{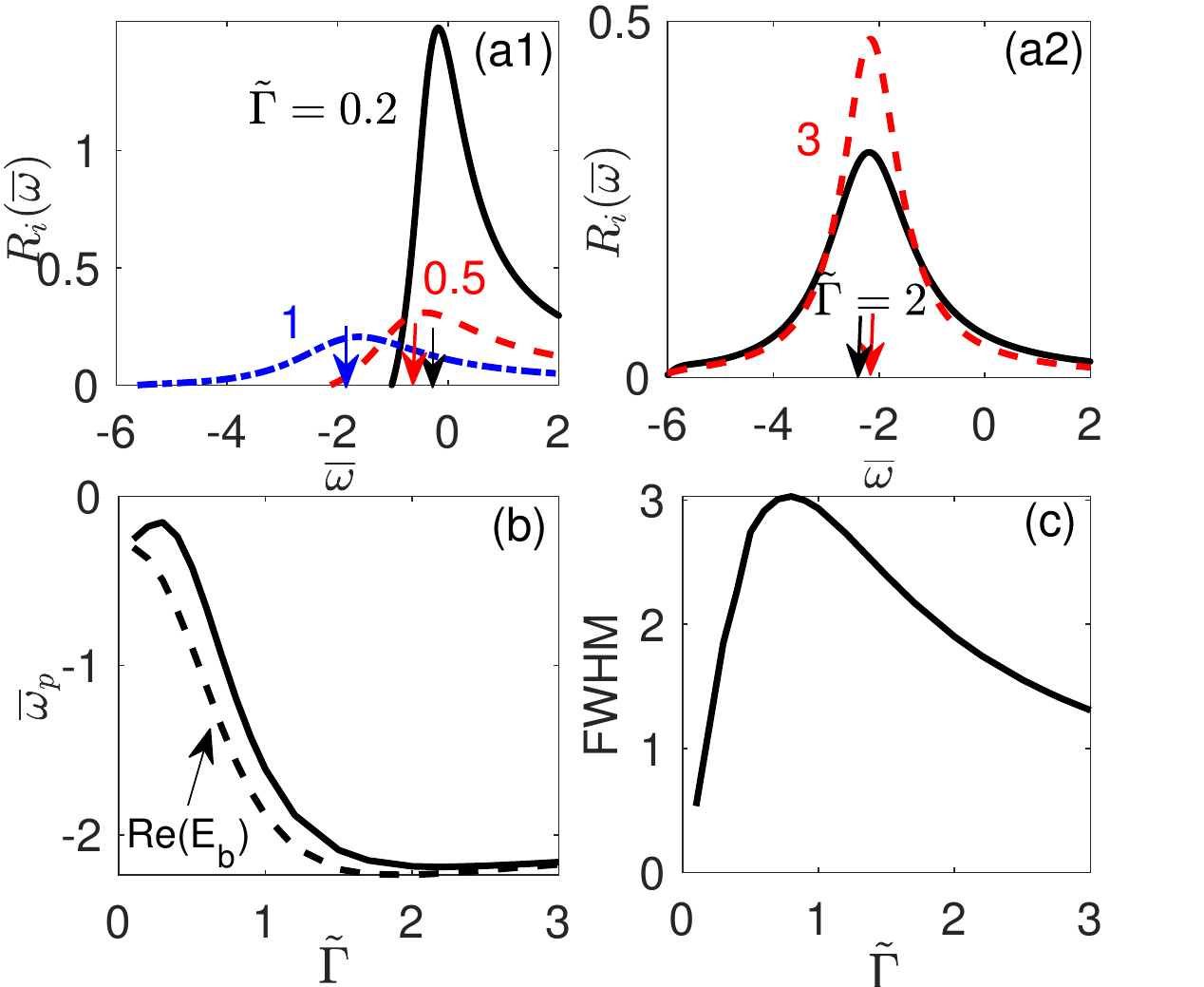}
\caption{(Color online) Molecule spectrum via the inverse r.f. spectroscopy under blue-detuned Raman lasers.
(a1)(a2) Transition rate $R_i$ at different $\tilde{\Gamma}$. We use the shifted frequency $\bar{\omega}$ (see main text) as $x$-axis, and mark the locations of $\text{Re}(E_b)$ for each case with arrows of the same color.
(b)(c) The peak position $\bar{\omega}_p$ and the FWHM of r.f. spectrum as functions of $\tilde{\Gamma}$. In (b) we also show $\text{Re}(E_b)$ to guide the eye. In all plots we take $\Omega_0=2$ and $1/a_s=1$. We have taken $q$ and $E_q$ as the units of momentum and energy, respectively.
} \label{fig_rf}
\end{figure}

In the present case, the comparison of $\text{Im}(E_b)$ and $\text{Im}(\xi_{\cp k \lambda})$ suggests that the requirement in (i) can hardly be satisfied, while in (ii) can be met over a considerable parameter range given the Raman laser is blue-detuned.
Indeed, the full numerical calculations of Eqs.~(\ref{rate_direct})(\ref{rate_inverse}) confirm the feasibility of achieving a positive transition rate using the inverse r.f. spectroscopy with blue-detuned Raman lasers. This is demonstrated in Figs.~\ref{fig_rf}(a1)(a2), where $R_i$ stays positive over a large frequency range, especially around the peak position $\bar{\omega}_p$. Note that we have used a shifted frequency $\bar{\omega}=\omega-2\xi_{\rm th}+\xi_{q-}$ in Fig.~\ref{fig_rf}, for a better comparison of  $\bar{\omega}_p$ and $\text{Re}(E_b)$.

In Fig.~\ref{fig_rf}, the evolution of $R_i(\bar{\omega})$ with $\tilde{\Gamma}$ can be classified into two regimes. When $\tilde{\Gamma}$ is small [Fig.\ref{fig_rf}(a1)], increasing $\tilde{\Gamma}$ leads to a spectral broadening, with the peak location $\bar{\omega}_p$  getting more negative, consistent with a similar trend in $\text{Re}(E_b)$. This is the regime where the molecule formation is greatly enhanced by dissipation. On the other hand, when $\tilde{\Gamma}$ is sufficiently large [Fig.\ref{fig_rf}(a2)], further increase in $\tilde{\Gamma}$ leads to a narrower spectral width, with $\bar{\omega}_p$ getting less negative. This is the regime where the molecule is less affected by non-Hermitian SOC due to the quantum Zeno effect. In Figs.~\ref{fig_rf}(b)(c), we extract the peak position $\bar{\omega}_p$ and the full width at half magnitude (FWHM) of the r.f. spectrum, both of which show non-monotonic evolutions with increasing $\tilde{\Gamma}$. In particular, the evolution of $\bar{\omega}_p$ indeed follows the same trend as in $\text{Re}(E_b)$ [Fig.\ref{fig_rf}(b)].  We therefore conclude that the r.f. spectrum is experimentally detectable and captures the key properties of dissipation-facilitated molecules.

\section{Discussion}
In summary, we have proposed a realistic experimental scheme to explore the interplay of interaction and non-Hermitian SOC. We demonstrate the existence of dissipation-facilitated molecules, and show that they can be detected using inverse r.f. spectroscopy.
Our model (\ref{h0_single}) can be applied to a variety of atomic species that are actively studied in cold atoms. An explicit example is the $^{40}$K Fermi gas, where r.f. spectrum has been successfully measured under the Hermitian SOC, both without~\cite{SOC_expt3} and with~\cite{Zhang_expt} interactions. For instance, the two hyperfine states are chosen as $\{|F=\frac{9}{2},-\frac{7}{2}\rangle,|F=\frac{9}{2},-\frac{9}{2}\rangle\}$ in the $^2S_{1/2}$ manifold~\cite{Zhang_expt}, and an additional dressing laser can be applied to the excited $^2P_{1/2}$ manifold for tunable dissipation. Under the parameters $\delta\sim 50$MHz, $\Gamma\sim 100$MHz, and $\eta\sim 300$kHz, we have $\gamma=2\text{Im}(\Omega)\sim 1.8$kHz, which suggests that our model should be valid for $t\ll 0.55$ms. Since $\tau_m\sim 1/\gamma$, the lifetime of the dissipative molecule is also on the order of $0.55$ms, making the molecules stable enough to be detected for $t\ll 0.55$ms.

Furthermore, we note that the validity condition of $H_0$ and $H_{\rm SOC}$ should be separated from the ability to detect the molecules. The former requires the measurement be completed within a short time ($t\ll 1/\gamma=1/|2{\rm Im}(\Omega)|$), while the latter would require a comparison of  lifetimes between the initial and final states in the r.f. spectroscopy. Both requirements need to be satisfied in practical experiments in order to detect the molecular physics discussed in this work.
Moreover, the timescale over which the system relaxes to the quasi-steady molecular state actually sensitively depends on the wave-function overlap between the initial state and the quasi-steady molecular state. In general, the few-body relaxation time is much shorter than that is required for the many-body correlated state. As the limit of short-time scale ($1/|2{\rm Im}(\Omega)|$) can be manipulated to be long in experiment (for instance, through a small $\Omega_0$ generated by a smaller Raman intensity and larger detuning), this time can be sufficiently long for the stabilization of the quasi-steady molecular state.

Our results are expected to serve as a guide for the experimental observation of non-trivial effects of non-Hermiticity in interacting Fermi gases, which has so far eluded experimental efforts. Future studies would include the dissipation-induced Fermi superfluidity in the corresponding many-body systems, as well as other exotic few- and many-body quantum states where non-Hermiticity plays a key role.

\acknowledgements
The work is supported by the National Key Research and Development Program of China (2018YFA0307600, 2016YFA0300603, 2016YFA0301700, 2017YFA0304100),  the National Natural Science Foundation of China (No.11534014, No.11974331) and the Strategic Priority Research Program of Chinese Academy of Sciences (No. XDB33000000).

\appendix
\section{Perturbative expansion} \label{app_A}

The non-monotonic behavior of $k_{\rm min}$ and $C$ can be understood through perturbation theory in either the small- or large-$\tilde{\Gamma}$ limit. Here $k_{\rm min}$ is the location of minimum $\text{Re}(\xi_{\cp k -})$ along the $x$ direction, and $C$ is defined in Eq.~(\ref{C}).

For $\tilde{\Gamma}\ll 1$, the dissipation can be treated as perturbation. Up to the lowest order, we have $\Omega=\Omega_0(1+i\tilde{\Gamma})$, whose imaginary part $\text{Im}(\Omega)\propto \tilde{\Gamma}$. This leads to the analytical expressions $k_{\rm min}=q\sqrt{1-s^2(1-3\tilde{\Gamma}^2)}$ and $C=1-s^2(1-3\tilde{\Gamma}^2)$, with $s= \Omega_0/(2E_q)\ll 1$. Thus, both $k_{\rm min}$ and $C$ increase quadratically with $\tilde{\Gamma}$. In the opposite limit $\tilde{\Gamma}\gg 1$, however, the SOC strength decays as $\Omega\sim i\Omega_0/\tilde{\Gamma}$, and the $\Omega$ term behaves just like an imaginary magnetic field, i.e., the magnetic field strength is a pure imaginary number.
In this limit, we treat $\Omega$ as perturbation, and get $k_{\rm min}=q\sqrt{1+s^2/\tilde{\Gamma}^2}$ and $C=1+s^2/\tilde{\Gamma}^2$, both decreasing with larger $\tilde{\Gamma}$. The distinct trends of $k_{\rm min}$ and $C$ in different limits of $\tilde{\Gamma}$ explain the non-monotonicity shown in Fig.~1 of the main text. We also note that behavior in the large $\tilde{\Gamma}$ limit can be understood through the quantum Zeno effects, which effectively suppress the off-diagonal SOC term.

\section{Fermi's golden rule modified by non-Hermiticity} \label{app_B}

In the Hermitian limit with $\tilde{\Gamma}=0$, Eqs.~(31,32) are reduced to the familiar Fermi's golden rule of the corresponding r.f. process in spin-orbit-coupled Hermitian systems, with experimentally detectable positive transfer rate $R(\omega)>0$. Specifically, in the Hermitian limit, $|{\cp k}^R_{\lambda}\rangle=|{\cp k}^L_{\lambda}\rangle\equiv |{\cp k}_{\lambda}\rangle$, $|\Phi^R\rangle=|\Phi^L\rangle\equiv|\Phi\rangle$, and all energies are real. It follows that Eqs.~(31,32) exactly reproduce the formula of Fermi's golden rule under the corresponding r.f. process. For instance, the direct r.f. spectrum is reduced from Eq.~(31) to
\begin{equation}
R_d(\omega)=\pi\sum_{\lambda=\pm} |\langle -{\cp k}_{\lambda}|\bar{\psi}_{{\cp k} \downarrow}|\Phi\rangle|^2\delta(\omega+E_2-\xi_{-{\cp k}\lambda}-\epsilon_{{\cp k}+q{\cp e}_x}).  \label{rate2}
\end{equation}
The Fermi's golden rule results in a positive-definite transition rate as studied earlier in SOC systems~\cite{Hu,Zhang_expt}. However, for the non-Hermitian case here, both the numerator and denominator in the summand of Eqs.~(31,32) can be complex. The conventional Fermi's golden rule breaks down, and the resulting $R(\omega)$ is significantly modified and not positive-definite anymore.


\begin{thebibliography}{99}
\bibitem{footnote}The dissipation therein effectively generates a magnetic field strength that is a pure imaginary number. Such field produces zero eigenvalue when act on the spin-singlet state.
\bibitem{footnote_helicity} The requirement is more stringent for fermions in the lower-helicity branch with $\lambda=-$ in Eqs.~(\ref{rate_direct})(\ref{rate_inverse}), which contribute the largest weight in the molecular wavefunction. For upper-helicity fermions with $\lambda=+$, the requirement can be relaxed.
\end{thebibliography}
\end{document}